

\input phyzzx

\def\p{\partial}
\def\loo{\lambda^{(0)}_1}
\def\lot{\lambda^{(0)}_2}
\def\mo{M^{(0)}}
\def\lo{\lambda^{(0)}}

{}~\hfill\vbox{\hbox{TIFR/TH/92-46}\hbox{hepth@xxx/9209016}
\hbox{September, 1992}}\break

\title{QUANTIZATION OF DYON CHARGE AND ELECTRIC-MAGNETIC DUALITY IN
STRING THEORY}

\author{Ashoke Sen}

\address{Tata Institute of Fundamental Research, Homi Bhabha Road, Bombay
400005, India}

\let\refmark=\NPrefmark 
\def\define#1#2\par{\def#1{\Ref#1{#2}\edef#1{\noexpand\refmark{#1}}}}
\def\con#1#2\noc{\let\?=\Ref\let\<=\refmark\let\Ref=\REFS
         \let\refmark=\undefined#1\let\Ref=\REFSCON#2
         \let\Ref=\?\let\refmark=\<\refsend}

\define\RSTRING
A. Sen, preprint TIFR-TH-92-39 (hepth@xxx/9206016).

\define\RNARAIN
K. Narain, Phys. Lett. {\bf B169} (1986) 41.

\define\RNSW
K. Narain, H. Sarmadi and E. Witten, Nucl. Phys. {\bf B279} (1987) 369.

\define\RORT
T. Ortin, preprint SU-ITP-92-24 (hepth@xxx/9208078).

\define\RDGHR
A. Dabholkar, G. Gibbons, J. Harvey and F.R. Ruiz, Nucl. Phys. {\bf B340}
(1990) 33;
A. Dabholkar and J. Harvey, Phys. Rev. Lett. {\bf 63} (1989) 719.

\define\ROLIVE
C. Montonen and D. Olive, Phys. Lett. {\bf B72} (1977) 117;
H. Osborn, Phys. Lett. {\bf B83} (1979) 321.

\define\RIBANEZ
A. Font, L. Ibanez, D. Lust and F. Quevedo, Phys. Lett. {\bf B249} (1990)
35;
J. Harvey and J. Liu, Phys. Lett. {\bf B268} (1991) 40;
S.J. Rey, Phys. Rev.{\bf D43} (1991) 526.

\define\RSELF
J. Cardy, Nucl. Phys. {\bf B205} (1982) 17;
A. Shapere and F. Wilczek, Nucl. Phys. {\bf B320} (1989) 669.

\define\RTYPE
M. Duff and J. Lu, Phys. Lett. {\bf B273} (1991) 409.

\define\RSTROM
A. Strominger, Nucl. Phys. {\bf B343} (1990) 167; C. Callan, J. Harvey and
A. Strominger, Nucl. Phys. {\bf B359} (1991) 611; {\bf B367} (1991) 60;
preprint EFI-91-66.

\define\RPES
I. Pesando and A. Tollsten, Phys. Lett. {\bf B274} (1992) 374.

\define\RDUFF
M. Duff, Class. Quantum Grav. {\bf 5} (1988) 189;
M. Duff and J. Lu, Nucl. Phys. {\bf B354} (1991) 129, 141; {\bf B357}
(1991) 354;
Phys. Rev. Lett. {\bf 66} (1991) 1402;
Class. Quantum Grav. {\bf 9} (1991) 1;
M. Duff, R. Khuri and J. Lu, Nucl. Phys. {\bf B377} (1992) 281.
J. Dixon, M. Duff and J. Plefka, preprint CTP-TAMU-60/92
(hepth@xxx/9208055).

\define\RGAZU
M. Gaillard and B. Zumino, Nucl. Phys. {\bf B193} (1981) 221.

\define\RODD
S. Ferrara, J. Scherk and B. Zumino, Nucl. Phys. {\bf B121} (1977) 393;
E. Cremmer, J. Scherk and S. Ferrara, Phys. Lett. {\bf B68} (1977) 234;
{\bf B74} (1978) 61;
E. Cremmer and J. Scherk, Nucl. Phys. {\bf B127} (1977) 259;
E. Cremmer and B. Julia, Nucl. Phys.{\bf B159} (1979) 141;
M. De Roo, Nucl. Phys. {\bf B255} (1985) 515; Phys. Lett. {\bf B156}
(1985) 331;
E. Bergshoef, I.G. Koh and E. Sezgin, Phys. Lett. {\bf B155} (1985) 71;
M. De Roo and P. Wagemans, Nucl. Phys. {\bf B262} (1985) 646;
L. Castellani, A. Ceresole, S. Ferrara, R. D'Auria, P. Fre and E. Maina,
Nucl. Phys. {\bf B268} (1986) 317; Phys. Lett. {\bf B161} (1985) 91;
S. Cecotti, S. Ferrara and L. Girardello, Nucl. Phys. {\bf B308} (1988)
436;
M. Duff, Nucl. Phys. {\bf B335} (1990) 610.

\define\RMAHSCH
J. Maharana and J. Schwarz, preprint CALT-68-1790 (hepth@xxx/9207016).

\define\RTSW
A. Shapere, S. Trivedi and F. Wilczek, Mod. Phys. Lett. {\bf A6}
(1991) 2677.

\define\RSDUAL
A. Sen, preprint TIFR-TH-92-41 (hepth@xxx/9207053).

\define\RDIRAC
P. Dirac, Proc. R. Soc. {\bf A133} (1931) 60.

\define\RWITTENTHETA
E. Witten, Phys. Lett. {\bf 86B} (1979) 283.

\abstract

We analyze the allowed spectrum of electric and magnetic charges carried
by dyons in (toroidally compactified) heterotic string
theory in four dimensions at arbitrary values of the string coupling
constant and $\theta$ angle.
The spectrum is shown to be invariant under electric-magnetic duality
transformation, thereby providing support to the conjecture that this is
an exact symmetry in string theory.

\endpage

It has recently been shown\RTSW\RSDUAL\ that the equations of motion
derived from the low energy effective action in four dimensional string
theory are invariant under the electric-magnetic duality
transformation\RGAZU\RODD\ that interchanges the electric and magnetic
fields, and at the same time interchanges the strong and
weak coupling limits of the theory.
Using the fact that at least in the four dimensional string theory
obtained by toroidal compactification of ten dimensional heterotic string
theory, there exists a string like solution in this effective field theory
whose zero modes are in one to one correspondence to the dynamical
degrees of freedom of the fundamental heterotic string in four
dimensions\RDGHR\RSTRING, it was argued in ref.\RSDUAL\ that the effective
field theory contains all the information about the full string theory,
and hence the duality symmetry of the effective field theory might imply
duality symmetry of the full string theory under which electrically
charged particles get interchanged with magnetically charged particles.
Earlier conjectures to this effect in field theory was made in
refs.\ROLIVE, and in string theory in refs.\RIBANEZ.
Similar duality between ten dimensional heterotic string theory and
five-brane theory was conjectured in refs.\RSTROM\RDUFF\RPES.
Finally, appplications of this duality transformation to generate new
classical solutions in the effective field theory were made in
refs.\RTSW\RSDUAL\RORT.

Our analysis in ref.\RSDUAL\ has been purely classical.
In this paper we shall analyze the compatibility of the duality conjecture
with the well known quantization condition of the electric and magnetic
charges of a dyon\RDIRAC.
These conditions are known to receive non-trivial modifications in the
presence of the theta angle\RWITTENTHETA.
We shall show that the quantization rules are
invariant under duality transformation, i.e. if we start from a given
value of the string coupling constant $g$ and $\theta$-angle, and then
perform a duality transformation that changes $g$ to $g'$ and $\theta$ to
$\theta'$, then the transformed dyon state is one of the allowed states
for string coupling $g'$ and angle $\theta'$.
Related work for type IIB superstring theory in ten dimensions was carried
out in ref.\RTYPE.
Effect of $\theta$-terms on self dual laattice models was studied in
refs.\RSELF.

The result of refs.\RMAHSCH\RSDUAL\ may be summarized as follows.
The low energy effective field theory describing toroidally compactified
heterotic string theory in four dimensions contains the metric
$G_{\mu\nu}$, 28 vector fields
$A_\mu^{(\alpha)}$ ($1\le\alpha\le 28$), a complex scalar field
$\lambda=\lambda_1+i\lambda_2$,
and a set of scalar fields that can be described by a 28$\times$28 matrix
$M$, satisfying,
$$
M^T=M, ~~~~~~~ M^T L M=L
\eqn\eone
$$
where,
$$
L=\pmatrix{ 0 & I_6 & 0\cr I_6 & 0 & 0\cr 0 & 0 & -I_{16}\cr}
\eqn\etwo
$$
is a 28$\times$28 matrix.
$I_n$ denotes the $n\times n$ identity matrix.
The equations of motion of the low energy effective field theory follow
from the action:
$$\eqalign{
S =&{1\over 32\pi}\int d^4 x\sqrt{-\det G}[ R -{1\over 2(\lambda_2)^2}
G^{\mu\nu}\p_\mu\lambda\p_\mu\bar\lambda -\lambda_2\vec F^T_{\mu\nu}.LML.
\vec F^{\mu\nu} \cr
& +\lambda_1\vec F^T_{\mu\nu}.L.\vec{\tilde F}^{\mu\nu} +{1\over 8}
G^{\mu\nu}
Tr(\p_\mu M L\p_\nu M L)]\cr
}
\eqn\ethree
$$
Here the arrow on $\vec F_{\mu\nu}$ denotes that it is a 28 dimensional
vector:
$$
F^{(\alpha)}_{\mu\nu} =\p_\mu A^{(\alpha)}_\nu - \p_\nu A^{(\alpha)}_\mu
\eqn\efour
$$
$\tilde F^{(\alpha)}_{\mu\nu}$ denotes the dual of
$F^{(\alpha)}_{\mu\nu}$.
The vector $\vec F_{\mu\nu}$ is related to a similar vector defined in
ref.\RSDUAL\ by a factor of 2.
Also $S$ has been multiplied by an overall factor of $1/32\pi$, which does
not affect the classical equations of  motion and hence the analysis of
ref.\RSDUAL.\foot{This factor can always be absorbed into a rescaling
of $\lambda$ and $G_{\mu\nu}$.}
The equations of motion derived from this action (but not the action
itself) are invariant under the following two transformations:
$$
\lambda\to\lambda +1, ~~~~ \vec F_{\mu\nu}\to \vec F_{\mu\nu}, ~~~~ M\to
M, ~~~~ G_{\mu\nu}\to G_{\mu\nu}
\eqn\efive
$$
and,
$$\eqalign{
& \lambda\to\lambda' = -{1\over\lambda}, ~~~~ \vec F_{\mu\nu}\to \vec
F'_{\mu\nu} = -\lambda_2 ML\vec{\tilde F}_{\mu\nu} - \lambda_1 \vec
F_{\mu\nu}\cr
& M\to M, ~~~~ G_{\mu\nu}\to G_{\mu\nu}\cr
}
\eqn\esix
$$
Together these generate an SL(2,{\bf Z}) symmmetry.
Although the equations of motion are invariant under $\lambda\to\lambda
+c$ for any real number $c$, this symmetry is broken down to $\lambda\to
\lambda +1$ by instanton corrections\RTSW.

Since it helps us to fix some of the as yet undetermined normalizations in
the theory, we shall now show how to obtain the above result.
Let us consider a specific embedding of one of the U(1) gauge fields (say
$A_\mu^{(28)}$) into an SU(2) subgroup of one of the $E_8$ (or SO(32)/{\bf
Z$_2$}) groups of the heterotic string theory.
Using the  freedom of scaling $\lambda$ by a constant $c$ and $\vec F$ by
$1/\sqrt c$ which leaves the action invariant, we can always ensure that
$A_\mu^{(28)}$ is related to the third component $B_{3\mu}$ of the SU(2)
gauge fields as $B_{3\mu}=A_\mu^{(28)}/\sqrt 2$.\foot{The factor of $\sqrt
2$ also provides a normalization such that the electric
charge vector, defined later, takes value on an even, self-dual lattice.}
The term proportional to $F^{(28)}_{\mu\nu}\tilde F^{(28)\mu\nu}$ in the
action then
becomes part of an SU(2) invariant term,
$$
{1\over 16\pi} \int d^4 x \sqrt{-\det G} ~ \lambda_1 \sum_{a=1}^3
H_{a\mu\nu} \tilde H_a^{\mu\nu}
\eqn\esixa
$$
where $H_{a\mu\nu}=\p_\mu B_{a\nu} -\p_\nu B_{a\mu} +\epsilon^{abc}
B_{b\mu} B_{c\nu}$.
Using the relation,
$$
\int d^4 x \sqrt{-\det G}  \sum_{a=1}^3
H_{a\mu\nu} \tilde H_a^{\mu\nu} = 32\pi^2
\eqn\esixb
$$
for a single instanton configuration, we see that $S$ changes by $2\pi$
times an integer under $\lambda_1\to\lambda_1 +1$, and hence $e^{iS}$
remains invariant under this transformation.
Symmetry under a general transformation of the form $\lambda\to\lambda +c$
is broken by the instanton corrections.

Let us now denote by $\lo=\loo+i\lot$ the asymptotic value of $\lambda$,
and by $\mo$ the asymptotic value of the matrix $M$.
{}From the form of the action we see that $\loo$ and $\lot$ are related to
the string coupling constant $g$ and the $\theta$ angle by the relations:
$$
\lot = {8\pi\over g^2}, ~~~~~ \loo ={\theta\over 2\pi}
\eqn\eseven
$$
Finally, for a given state, we define the 28 dimensional electric and
magnetic charge vectors $\vec Q_e$ and $\vec Q_m$ in terms of the
asymptotic form of $\vec F_{\mu\nu}$ as follows:
$$
F^{(\alpha)}_{0 r} \simeq {Q_e^{(\alpha)}\over r^2}, ~~~~~
\tilde F^{(\alpha)}_{0 r} \simeq {Q_m^{(\alpha)}\over r^2}
\eqn\eeight
$$
{}From eqs.\esix\ and \eeight\ we see that under a duality
transformation,
$$\eqalign{
&\loo\to {\loo}'= -{\loo\over |\lo|^2}, ~~~~
\lot\to {\lot}' = {\lot\over |\lo|^2}\cr
&\vec Q_e\to \vec Q_e' = -\lot\mo L\vec Q_m -\loo \vec Q_e\cr
&\vec Q_m\to\vec Q_m' = \lot\mo L\vec Q_e -\loo\vec Q_m\cr
}
\eqn\enine
$$
Let us now study the spectrum $(\vec Q_m, \vec Q_e)$ of magnetic and
electric  charges in this theory.
We start from the states with zero magnetic charge.
With the normalization convention that we have adopted, the charge
spectrum of such states  is given by\RNARAIN\RNSW,
$$
(\vec Q_m, \vec Q_e) = (0, {1\over\lot}\vec\alpha)
\eqn\eeleven
$$
where $\vec\alpha$ is a lattice vector belonging to a 28 dimensional
self-dual, even, Lorentzian lattice with metric $L$.
Let us denote this lattice by $P$.
Then, for $\vec\alpha, \vec\beta\in P$, we have,
$$
\vec\alpha^T . L . \vec \beta = integer, ~~~~
\vec\alpha^T . L . \vec\alpha = even ~ integer
\eqn\eten
$$

Let us now consider a general dyon state $(\vec Q_m, \vec Q_e)$.
A consistent spectrum of $\vec Q_m$ is obtained by demanding that the
Dirac string attached to the magnetic charge is not visible to the
particle of charge $(0, \vec\alpha/\lot)$.
With the normalization convention we have adopted, this requires,
$$
\lot \vec Q_m . L\mo L . {1\over \lot}\vec\alpha = integer
\eqn\etwelve
$$
Using eqs.\eone, \etwo, and the self duality of the lattice $P$ we see
that the most general solution of eq.\etwelve\ is of the form:
$$
\vec Q_m =\mo L\vec\beta, ~~~~~\vec \beta \in P
\eqn\ethirteen
$$
We now try to determine the allowed values of $\vec Q_e$ for the value of
$\vec Q_m$ given in eq.\ethirteen.
Naively one might have expected that these are given by $\vec\alpha/\lot$
with $\vec\alpha\in P$ as in eq.\eeleven.
However we know that in the presence of a theta angle the allowed electric
charges of a dyon are shifted by an amount proportional to the magnetic
charge\RWITTENTHETA.
The shift in this case can be computed following the procedure of
ref.\RWITTENTHETA\ and is given by $\loo\vec\beta/\lot$.
Thus the spectrum of electric and magnetic charges carried by the dyon is
given by,
$$
(\vec Q_m, \vec Q_e) = \Big(\mo L\vec \beta, {1\over\lot}(\vec\alpha
+\loo\vec\beta) \Big), ~~~~~ \vec\alpha, \vec\beta\in P
\eqn\efourteen
$$

We can now easily perform a duality transformation and compute $(\vec
Q_m', \vec Q_e')$ using eq.\enine.
The result is,
$$
(\vec Q_m', \vec Q_e') = \Big(\mo L \vec \alpha, {1\over{\lot}'}(-\vec\beta
+{\loo}'\vec\alpha )\Big)
\eqn\efifteen
$$
This shows that the spectrum given by eq.\efourteen\ is invariant under
the duality transformation, since the transformed spectrum \efifteen\ has
the same form as the original spectrum \efourteen, with $\vec\alpha$,
$\vec\beta$ transforming as,
$$
\vec\alpha\to\vec\alpha' =-\vec\beta, ~~~~ \vec\beta \to\vec\beta'
=\vec\alpha
\eqn\esixteen
$$
Note also that under the other generator of the SL(2,{\bf Z})
transformation, $\loo\to {\loo}'=\loo +1$ with all other quantities
remaining fixed,
$$
(\vec Q_m', \vec Q_e') = (\vec Q_m, \vec Q_e) = \Big(\mo L\vec\beta,
{1\over {\lot}'} (\vec\alpha -\vec\beta +{\loo}'\vec\beta)\Big)
\eqn\eseventeen
$$
Thus the spectrum again retains its form, with the transformations,
$$
\vec\alpha\to\vec\alpha' = \vec\alpha -\vec\beta,
{}~~~~ \vec\beta\to \vec\beta' = \vec\beta
\eqn\eeighteen
$$

This establishes that the electric and the magnetic charge spectrum of
dyons is invariant under the full SL(2, {\bf Z}) group of transformations.
This, in turn, shows that the laws of quantization of dyon charge
are consistent with the idea that electric-magnetic duality is an exact
symmetry of four dimensional string theory.

\refout
\end